\documentclass[aps,prl,showpacs,footinbib,superscriptaddress,twocolumn,longbibliography]{revtex4-2}
\usepackage{amsfonts}
\usepackage{amsmath}
\usepackage{multirow}
\usepackage{txfonts}
\usepackage{amssymb}
\usepackage{amsbsy}
\usepackage{graphicx}
\usepackage{epstopdf}
\usepackage{color}
\usepackage{braket} 
\usepackage{mathdots} 
\usepackage{booktabs}
\usepackage{indentfirst}
\usepackage{hyperref}
\usepackage{diagbox}
\usepackage{lipsum}
\hypersetup{ colorlinks=true, linkcolor=blue, anchorcolor=blue, citecolor=blue,urlcolor=blue}

\begin{document}
\title{ Many-body non-Hermitian skin effect with exact steady states in the dissipative quantum link model}

\author{Yu-Min Hu}

\affiliation{Institute for Advanced Study, Tsinghua University, Beijing,  100084, China}
\affiliation{Max Planck Institute for the Physics of Complex Systems, N\"{o}thnitzer Str. 38, 01187 Dresden, Germany}
\author{Zijian Wang}
\affiliation{Institute for Advanced Study, Tsinghua University, Beijing,  100084, China}

\author{Biao Lian}
\altaffiliation{biao@princeton.edu}
\affiliation{Department of Physics, Princeton University, Princeton, New Jersey 08544, USA}

\author{Zhong Wang}
\altaffiliation{wangzhongemail@tsinghua.edu.cn}
\affiliation{Institute for Advanced Study, Tsinghua University, Beijing,  100084, China}  

\begin{abstract}
We introduce a dissipative lattice gauge model that exhibits the many-body version of the non-Hermitian skin effect. The dissipative couplings between dynamical gauge fields on the lattice links and the surrounding environment generate chiral motions of particles residing on lattice sites. Despite the complexity arising from many-body interactions, the local gauge symmetry enables the exact construction of a steady state that displays the many-body non-Hermitian skin effect. Furthermore, our approach can be generalized to realize a new type of many-body non-Hermitian skin effect, dubbed the hierarchical skin effect, where different subsystem degrees of freedom exhibit boundary accumulation of multiple moments at different orders. Our findings can be readily observed by engineering dissipation in state-of-the-art lattice gauge simulators.
\end{abstract}

\maketitle

\emph{Introduction.--}Gauge theory is a cornerstone in modern physics. The lattice formulation of gauge theory provides a versatile framework for investigating fundamental phenomena in quantum field theories and describing novel quantum phases in condensed matter physics \cite{Kogut1979, Kogut1983, fradkin2013field}. While it is difficult to simulate a strongly coupled gauge theory on classical computers, the rapid advancement of quantum technologies allows for experimentally investigating equilibrium and nonequilibrium phases of lattice gauge theory on cutting-edge quantum platforms \cite{wiese2013ultracold, Hauke2013quantum, Marcos2013superconducting, zohar2015quantum, martinez2016real, mil2020scalable, yang2020observation, zhou2022thermalization, cheng2024emergent,gyawali2024observation,busnaina2025native}. These novel phenomena includes string dynamics \cite{Banerjee2012atomic, Pichler2016realtime, Verdel2020realtime, Surace2020LGT, zhang2023observation,de2024observationstringbreakingdynamicsquantum,cochran2024visualizing}, quantum many-body scars \cite{bernien2017probing, Desaules2023Prominent, Desaules2023weak, Su2023observation}, disorder-free localization \cite{Smith2017disorderfree1,Karpov2021disorderfree,Halimeh2022disorderfree,Chakraborty2023disorderfree}, and quantum phase transitions \cite{Rico2014tensor, Huang2019dynamical, Yao2022quantum, Wang2023interrelated}. A key ingredient for lattice gauge simulators is to impose gauge constraints for quantum dynamics \cite{Stannigel2014Constrained, Halimeh2020reliability, Halimeh2020Robustness, Halimeh2021gauge}.  One of the gauge violation mechanisms stems from the incoherent system-environment couplings of quantum devices \cite{Stannigel2014Constrained}. Although the environment is detrimental to gauge-invariant coherent dynamics, the interplay between dissipation and many-body dynamics can lead to unexpected nonequilibrium phenomena that has no analog in closed lattice gauge theory~\cite{wang2023topologically,dai2023steadystate}.

One of the striking phenomena arising in non-Hermitian or open quantum systems is the non-Hermitian skin effect (NHSE), which describes the boundary localization of bulk eigenstates \cite{yao2018edge,lee2019anatomy,kunst2018biorthogonal, Ghatak2019NHSE, xiao2020non, helbig2020generalized, Weidemann2020topological}. This concept is essential in various aspects of non-Hermitian physics \cite{Bergholtz2021RMP, Ashida2021, Zhang2022ReivewOnNHSE,ding2022non, Lin_2023,okuma2023non}, such as non-Hermitian topology \cite{yao2018chern, Song2019real, Yokomizo2019non, Okuma2020topological, zhang2020correspondence, Yang2020non, Wang2024Amoeba} and novel dynamics \cite{song2019chiral, Longhi2019probing, Wanjura2019, xue2021simple, Xiao2021observation, Haga2021Liouvillian, Longhi2022healing, Xue2022non,xiao2023observation, Zhu2024Observation, Hu2024Geometric}. Recently, there has been a growing interest in exploring the significant role of interactions in shaping NHSE in many-body systems. Considerable efforts have been dedicated to characterizing the NHSE of many-body eigenstates \cite{Lee2020manybody,Mu2020Emergent, Alsallom2022Fate, Kawabata2022manybody, shimomura2024general, hamanaka2024multifractality}, engineering many-body NHSE \cite{lee2021manybody, Xu2021interaction, Faugno2022interaction, Hamanaka2023interaction, Poddubny2023interaction, Li2023manybody, shen2024enhanced, gliozzi2024manybody},  and unveiling novel nonequilibrium phases induced by many-body NHSE \cite{Kawabata2023entanglement, wang2023absence, hu2023steady, yoshida2024nonhermitian, Qin2024occupation, Begg2024quantum,qin2024dynamical,yang2024nonhermitianultrastrongbosoniccondensation,wang2024nonbloch}. 

However, apart from a limited number of integrable systems \cite{Wang2023Scale-free, li2023nonbloch, Mao2023non, Zheng2024exact, mao2024liouvillian, ekman2024liouvillian}, the complexity of many-body interactions generally prevents the exact description of non-Hermitian many-body systems, thereby limiting the analysis of many-body NHSE to numerical methods. Therefore,  it is desirable to develop new analytical method to understand the interplay between NHSE and interaction. Moreover, the extensive studies of many-body NHSE are often rooted in the many-body extensions of the Hatano-Nelson model, where an imaginary background gauge field causes the nonreciprocal hoppings of particles \cite{Hatano1996localization,Hatano1997vortex}.  Compared to experimental platforms for single-particle NHSE  \cite{ Ghatak2019NHSE, xiao2020non, helbig2020generalized, Weidemann2020topological}, it is more challenging to implement imaginary gauge fields in quantum many-body experiments. More fundamentally, a physical explanation of imaginary gauge fields directly from a gauge-theoretic perspective is still lacking.

In this paper, we address these questions by proposing a dissipative extension of the quantum link model (QLM) in the U(1) lattice gauge theory \cite{chandrasekharan1997quantum, wiese2013ultracold, Banerjee2012atomic}. With the experimentally accessible dissipation processes acting on the dynamical gauge fields at lattice links, the dissipative QLM effectively induces nonreciprocal hoppings for matter fields at lattice sites, resulting in many-body NHSE under open boundary conditions (OBC). Despite the many-body complexity of this open quantum system, the local gauge symmetry of the quantum-link Hamiltonian enables the analytical construction of the \emph{exact} steady state, which exhibits the many-body NHSE characterized by the accumulation of the dipole moment of matter fields. Notably, not limited to dissipative QLM,  our symmetry-based method for constructing exact many-body eigenstates is broadly applicable to a wide class of open quantum many-body systems. Furthermore,  the dissipative QLM offers a general method for achieving a \emph{hierarchical skin effect}, characterized by multipole moments for certain subsystem degrees of freedom. From an experimental perspective, our model provides an efficient way to implement nonreciprocal hoppings in state-of-the-art lattice gauge simulators.

\emph{Dissipative quantum link model.--}We first review the basic notions of U(1) QLM \cite{Banerjee2012atomic}, particularly focusing on open boundary conditions. The QLM of (1+1)D quantum electrodynamics has a quantum-link Hamiltonian ${H}=J\sum_{n=1}^{L-1}({\psi}_n^\dagger {U}_{n,n+1}{\psi}_{n+1}+\text{h.c.})$. The lattice system has $L$ sites and $L-1$ links. $\psi_n^\dagger$ and $ \psi_n$ are the fermionic creation and annihilation operators of matter fields at sites. $ U_{n,n+1}$ is the link operator that represents the U(1) gauge degree of freedom, conjugating to the electric field $ {E}_{n,n+1}$ with the relations $[{E}_{n, n+1},  {U}_{n, n+1}]={U}_{n, n+1}$ and $[{E}_{n, n+1},  {U}_{n, n+1}^\dagger]=-{U}_{n, n+1}^\dagger$.  The generators of the local U(1) gauge symmetry in QLM are given by $ G_n= \psi_n^\dagger \psi_n-( E_{n,n+1}- E_{n-1,n})$ for $1<n<L$. In addition, we have $ G_1= \psi_1^\dagger \psi_1- E_{1,2}$ and $ G_L= \psi_L^\dagger \psi_L+ E_{L-1,L}$ at the boundary sites. A gauge fixing condition $( G_{n}+g_n)\ket{\Phi}=0$ with background charges $g_n$ plays the role of the Gauss law in electrodynamics. In practical quantum simulations, the quantum links are replaced by spin operators with finite Hilbert space dimensions: $ U_{n,n+1}\to s^{+}_{n,n+1},\  U_{n,n+1}^\dagger\to s^{-}_{n,n+1},\  E_{n,n+1}\to s^{z}_{n,n+1}$. Additionally, we perform the Jordan-Wigner transformation and employ spin-${1}/{2}$ operators $\tau_n^{x,y,z}$ to describe fermionic operators on lattice sites. The resulting spin version of QLM becomes
\begin{equation}\label{eq:hamiltonian}
    {H}=J\sum_{n=1}^{L-1}({\tau}_n^+ {s}_{n,n+1}^+{\tau}_{n+1}^-+\text{h.c.}).
\end{equation}
With $\psi_n^\dagger \psi_n={\tau}_n^z+1/2$, the occupied (unoccupied) states at lattice sites are described by the eigenvalue of $ \tau_n^z$ being $1/2$ ($-1/2$). The local gauge generator is equivalent to $ G_n={\tau}_n^z-( s_{n,n+1}^z- s_{n-1,n}^z)$ for $1<n<L$, with boundary corrections $  G_1={\tau}_1^z- s_{1,2}^z$ and $ G_L={\tau}_L^z+ s_{L-1,L}^z$. Additionally, the total particle number $ N=\sum_n\psi_n^\dagger \psi_n=\sum_n( {\tau}_n^z+1/2)$ of matter fields is also conserved because of the global U(1) symmetry. Without loss of generality, we focus on the simplest $s=1/2$ case of link spins, with the basis vectors $\ket{\uparrow}$ and $\ket{\downarrow}$ on each link. Our results are also applicable in higher-spin cases.

\begin{figure}
    \centering
    \includegraphics[width=8.5cm]{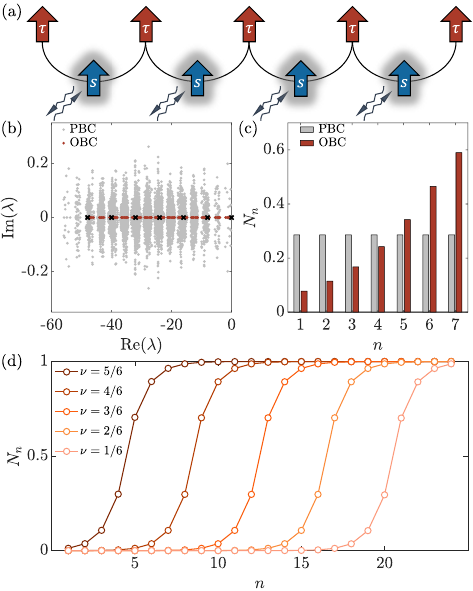}
    \caption{(a) Dissipative quantum link model with $ H$ in Eq.\eqref{eq:hamiltonian}.  (b) Liouvillian spectrums and (c) steady-state density distributions $N_n=\operatorname{Tr}[\rho_{\text{ss},N}( \tau_n+1/2)]$ numerically obtained by the exact diagonalization in the sector $\mathcal{G}_n=0$ and $N=2$ with parameters $L=7$, $J=1$, $\gamma_u=2.4$, and $\gamma_d=1.6$. We consider both the periodic (gray color) and open (red color) boundary conditions. Black points in (b) mark the eigenvalues of exact OBC eigenoperators. (d) The asymmetric particle distributions obtained from the exact OBC steady state.  $\nu=N/L$ is the filling factor. We take $\beta=\gamma_u/\gamma_d=3$ and $L=24$. }
    \label{fig:spectrum}
\end{figure}

We are interested in coupling the quantum links with a Markov bath [Fig.\ref{fig:spectrum}(a)], which is described by the Lindblad master equation:
\begin{equation}\label{eq:lindblad}
    \frac{\mathrm{d}}{\mathrm{d}t} \rho=\mathcal{L}[ \rho]=-i[ H, \rho]+\sum_{\mu}\left(2 L_\mu \rho  L_\mu^\dagger-\{ L_\mu^\dagger  L_\mu, \rho\}\right).
\end{equation}
The Liouvillian superoperator $\mathcal{L}$ defined in the double Hilbert space is divided into two parts $\mathcal{L}=\mathcal{L}_H+\mathcal{L}_D$, where $\mathcal{L}_H[\rho]=-i[ H, \rho]$ describes the coherent dynamics generated by the Hamiltonian $ H$ in Eq. \eqref{eq:hamiltonian} and $\mathcal{L}_D[\rho]=\sum_{\mu}(2 L_\mu \rho  L_\mu^\dagger-\{ L_\mu^\dagger  L_\mu, \rho\})$ denotes the dissipative dynamics associated with jump operators $ L_\mu$.

In the dissipative QLM, we couple each quantum link to the environment through two quantum jump operators describing biased incoherent spin-flipping processes:
\begin{equation}\label{eq:biased_jump}
     L_{n,n+1}^{(u)}=\sqrt{\gamma_u}{s}_{n,n+1}^{+} ,\quad  L_{n,n+1}^{(d)}=\sqrt{\gamma_d} {s}_{n,n+1}^{-}.
\end{equation}
These jump operators break the strong U(1) gauge symmetry generated by $ G_n$ into a \emph{weak gauge symmetry}, whose generator $\mathcal{G}_n$ acts on the operators $ O$ in the Hilbert space as $\mathcal{G}_n[ O]= G_n O- O G_n$. Therefore, we get $[\mathcal{G}_n,\mathcal{L}]=0$. In addition, the global charge $ N$ generates a strong symmetry of $\mathcal{L}$, satisfying $[ N, H]=[ N, L_\mu]=0$.

These strong and weak symmetries allow for using exact diagonalization to study the many-body spectrums of $\mathcal{L}$. We focus on the weak gauge sector $\mathcal{G}_n[\rho]=0$, to which the steady states belong.  If we add strong dissipators $ L_{n}^{(\text{gauge})}=\sqrt{\Gamma} G_n$ into Eq.\eqref{eq:lindblad} as an effective gauge fixing term \cite{Stannigel2014Constrained}, the relaxation dynamics will quickly converge to the above weak gauge sector. Nevertheless, these dissipators do not affect steady states and are thus not included in our construction. We use exact diagonalization to numerically obtain the spectrums of $\mathcal{L}$ under different boundary conditions\footnote{Under periodic boundary conditions, we add a link spin between the first site and the last site. Then the gauge generators are modified as $ G_1= {\tau}_1^z-( s_{1,2}^z- s_{L,1}^z)$ and $ G_L= \psi_L^\dagger {\tau}_L^z-( s_{L,1}^z- s_{L-1,L}^z)$. A similar modification is applied to $\mathcal{G}_{1}$ and $\mathcal{G}_{L}$. The periodic spectrum in Fig. \ref{fig:spectrum}(b) is numerically obtained by the exact diagonalization in the weak gauge sector $\mathcal{G}_n[\rho]=0$.}. As shown in Fig. \ref{fig:spectrum}(b), the OBC spectrum is enclosed by the spectrum under periodic boundary conditions (PBC) , which demonstrates the sensitivity of Liouvillian spectrums to boundary conditions \cite{supp_LGT}. Additionally, the steady state $\rho_{\text{ss}}$, satisfying $\mathcal{L}[\rho_{\text{ss}}]=0$, shows an asymmetric OBC distribution for the particle number $N_n=\operatorname{Tr}[\rho_{\text{ss}}( \tau_n+1/2)]$ [Fig.\ref{fig:spectrum}(c)]. These numerical results reveal the existence of many-body NHSE.

The emergence of NHSE can be understood as follows. First,
biased spin-flipping processes result in a steady state $\rho_{\text{ss}}$ with a preferable direction for link spins. For example, when $J=0$, $\braket{{s}_{n,n+1}^z}=\operatorname{Tr}[\rho_{\text{ss}} s_{n,n+1}^z]=(\beta-1)/(2\beta+2)$ depends on the ratio $\beta=\gamma_u/\gamma_d$. Intuitively, the dissipation-induced polarization of link spins can be statistically interpreted as a unidirectional electric field, rendering polarization-dependent interactions in $ H$ to transport particles in a favorable direction \cite{supp_LGT}.  This process effectively becomes nonreciprocal hoppings of matter fields, causing many-body NHSE. 

\emph{Strong dissipation limit.--}  In the strong dissipation limit $\gamma_{u,d}\gg J$, the emergent nonreciprocal hoppings of matter fields can be seen explicitly using perturbation theory \cite{Kessler2012generalized}. When $\mathcal{L}_{H}=0$, the unperturbed steady states of $\mathcal{L}_{D}$ are given by $\rho_{\text{ss}}^{(0)}=\rho_{\tau}\otimes\rho_{s}$. Here, $\rho_{\tau}$ is an arbitrary density matrix on lattice sites, and the steady-state density matrix on lattice links is 
\begin{equation}\label{eq:link_steady}
    \rho_{s}=\otimes_{n=1}^{L-1}\rho_{n,n+1},\quad \rho_{n,n+1}=\frac{\beta}{1+\beta}\ket{\uparrow}\bra{\uparrow}+\frac{1}{1+\beta}\ket{\downarrow}\bra{\downarrow}.
\end{equation}
Here, $\rho_{n,n+1}$ is defined on the $(n,n+1)$ link. 

A nonzero $\mathcal{L}_H$ perturbs these degenerate steady states,  and the second-order perturbation theory leads to an effective Liouvillian for $\rho_\tau$ of matter fields: $\mathcal{L}_{\text{eff}}[\rho_\tau]=\sum_\mu2 L_\mu\rho_\tau  L_\mu^\dagger-\{ L_\mu^\dagger  L_\mu,\rho_\tau\}$. The effective jump operators in $\mathcal{L}_{\text{eff}}$ are given by $ L_{n,n+1}^{(r)}=\sqrt{\gamma_r} \tau_n^{-} \tau_{n+1}^{+}$ and $ 
 L^{(l)}_{n,n+1}=\sqrt{\gamma_l} \tau_n^{+} \tau_{n+1}^{-}$, with $\gamma_r={ \gamma_u J^2}/{\left(\gamma_d+\gamma_u\right)^2}$ and $\gamma_l={\gamma_d J^2 }/{\left(\gamma_d+\gamma_u\right)^2}$ satisfying $\beta=\gamma_u/\gamma_d=\gamma_r/\gamma_l$.  Surprisingly, the effective Liouvillian $\mathcal{L}_{\text{eff}}$ describes a quantum asymmetric simple exclusion process, whose classical counterpart plays a significant role in statistical physics \cite{schutz1997exact,derrida1998exactly}. This effective model is exactly solvable based on the Bethe ansatz, exhibiting operator space fragmentation \cite{Essler2020integrability}. 

Therefore, dissipative QLM offers a practical approach for engineering nonreciprocal  hoppings for particles in a quantum many-body system, leading to unidirectional charge transport. Despite its similarity to the imaginary background gauge fields in the Hatano-Nelson model \cite{Hatano1996localization,Hatano1997vortex}, the gauge fields in our model are dynamically coupled to the environment in an experimentally accessible way.

\emph{Exact steady state.--} The emergent integrability discussed above only holds when $\gamma_u, \gamma_d \gg J$. It is generally a challenging task to determine the exact many-body Liouvillian spectrum. Nevertheless, the gauge structure of the quantum link Hamiltonian enables analytical construction of the OBC steady states, as well as certain other eigenstates, for arbitrary parameters. The construction directly demonstrates many-body NHSE in the dissipative quantum link model and, more interestingly, provides a general approach to obtaining exact steady states in a wide class of open quantum systems.

When $\mathcal{L}_H=0$,  the steady-state density matrix $\rho_s$ of $\mathcal{L}_D$ for link spins is a diagonal operator [Eq. \eqref{eq:link_steady}]. Following this observation, we introduce a double-space similarity transformation $\mathcal{T}=\mathcal{T}_\tau\otimes\mathcal{T}_s$. While $\mathcal{T}_\tau$ acting on sites will be specified shortly, $\mathcal{T}_s$ acting on links is designed to provide $\mathcal{T}_s[\rho_s]= T_s\rho_{s}= I_s$. Here, $ I_s$ is the identity operator in link Hilbert space, and $\rho_s$ is given by Eq. \eqref{eq:link_steady}. Consequently,  $ T_s=\otimes_{n=1}^{L-1} T_{n,n+1}$ with $ T_{n,n+1}=(1+\beta)/\beta\ket{\uparrow}\bra{\uparrow}+(1+\beta)\ket{\downarrow}\bra{\downarrow}\propto \exp(-\ln\beta s^z_{n,n+1})$ acting on the $(n,n+1)$ link \footnote{Due to the diagonal nature of $ \rho_s$ and $ T_s$, there is a redundancy in defining $\mathcal{T}_s$. For example, the form $\mathcal{T}_s[\rho_s]=\sqrt{ T_s}\rho_{s}\sqrt{ T_s}= I_s$ results in the same steady-state solution.}. It is easy to show that $\mathcal{T}_s\mathcal{L}_D\mathcal{T}^{-1}_s=\mathcal{L}_D^\dagger$ \footnote{Assume that $O$  is an arbitrary operator in the Hilbert space. $\mathcal{T}_s\mathcal{L}_D\mathcal{T}^{-1}_s$ acting on $O$ represents $\mathcal{T}_s\mathcal{L}_D\mathcal{T}^{-1}_s[O]=\mathcal{T}_s[\mathcal{L}_D[\mathcal{T}^{-1}_s[O]]]$. Additionally, the inverse of an invertible superoperator $\mathcal{S}$ is defined as $\mathcal{S}^{-1}\mathcal{S}[O]=\mathcal{S}\mathcal{S}^{-1}[O]=O$.} where $\mathcal{L}_D^\dagger[\rho]=\sum_{\mu}(2 L_\mu^\dagger \rho  L_\mu-\{ L_\mu^\dagger  L_\mu, \rho\})$ with $ L_\mu$ provided by Eq. \eqref{eq:biased_jump}. This result implies that $\mathcal{L}_D$ satisfies the quantum detailed balance condition \cite{alicki1976detailed,kossakowski1977quantum,fagnola2007generators,Firanko_2024}. Notably, the structure of the Liouvillian guarantees that $\mathcal{L}_D^\dagger[ I]=0$ with $ I$ being the identity matrix in the whole Hilbert space. A nonzero $\mathcal{L}_H$ also satisfies $\mathcal{L}_H[ I]=0$. Therefore, if there exists a double-space transformation $\mathcal{T}$ that keeps $\mathcal{L}_H$ invariant, i.e., $\mathcal{T}\mathcal{L}_H\mathcal{T}^{-1}=\mathcal{L}_H$, we can conclude that the identity operator $ I$ is the steady state of the deformed Liouvillian $\mathcal{T}\mathcal{L}\mathcal{T}^{-1}=\mathcal{L}_H+\mathcal{L}_D^\dagger$. Then the steady state of the original Liouvillian is given by $\rho_{\text{ss}}=\mathcal{T}^{-1}[ I]/\operatorname{Tr}(\mathcal{T}^{-1}[ I])$.  

The fact that the quantum-link Hamiltonian $ H$ satisfies the local gauge symmetry $[ G_n,  H]=0$ motivates us to construct $\mathcal{T}=\mathcal{T}_\tau\otimes\mathcal{T}_s$  by some generalized gauge transformation. Therefore, the  transformation $\mathcal{T}=\mathcal{T}_\tau\otimes\mathcal{T}_s$ compatible with $\mathcal{T}_s$ determined above is provided by $\mathcal{T}[ \rho]= T\rho$ where $ T=\exp[-\sum_{n=1}^{L}(\ln\alpha+n\ln\beta) G_n]$. Here $\alpha$ is an arbitrary constant. It is straightforward to show that $ T H T^{-1}= H$, and that $ T$ acting on links is equivalent to the above $ T_s$ up to an overall factor. In the end, we obtain the exact OBC steady state 
\begin{equation}\label{eq:exact_steady}
    \rho_{\text{ss}}=\frac{ T^{-1}}{\operatorname{Tr}( T^{-1})}=Z^{-1}\exp\left[\sum_{n=1}^{L}(\ln\alpha+n\ln\beta) G_n\right]
\end{equation}
with $Z=\operatorname{Tr}[\exp(\sum_{n=1}^{L}(\ln\alpha+n\ln\beta) G_n)]$.  This is one of the central results of this paper. We emphasize that the construction in Eq. \eqref{eq:exact_steady} is only valid under OBC.  The lack of translation symmetry makes Eq. \eqref{eq:exact_steady} incompatible with PBC. The mismatch between two boundary conditions is a signal of many-body NHSE. In the following, we give several remarks regarding this exact many-body steady state. 

Recall that the conserved global U(1) charge is $ N=\sum_{n=1}^L( \tau_n+1/2)$. If we define the global dipole charge $ D=\sum_{n=1}^L \tau_n[n-(L+1)/2]$ and the total $z$-component of link spins $ S^z=\sum_{n=1}^{L-1} s_{n,n+1}^z$, the steady state becomes $ \rho_{\text{ss}}\propto\exp[\ln\alpha N+\ln\beta( D+ S^z)]$. Because $ N$ is a conserved quantity, $\ln\alpha$ can be viewed as the chemical potential of a Gibbs ensemble. We can also obtain the $N$-particle steady state $ \rho_{\text{ss},N}\equiv Z^{-1}_N P_N\rho_{\text{ss}} P_N$, where $ P_N$ is the projection operator into the $N$-particle Hilbert space and $Z^{-1}_N$ is a normalization factor. Notably, $ \rho_{\text{ss},N}$ is independent of the free parameter $\alpha$. Therefore, the OBC steady states of Eq.\eqref{eq:lindblad} have an $(L+1)$-fold degeneracy.

When $0< N< L$,  the steady state $ \rho_{\text{ss},N}$ exhibits many-body NHSE by showing an asymmetric particle density at the lattice sites. As shown in Fig.\ref{fig:spectrum}(d), while the parameter $\beta$ shapes the localization behavior, the filling number $N/L$ determines the locus of the emergent real-space Fermi surface \cite{Mu2020Emergent}. The link spins, on the other hand, form a uniform configuration with $\braket{{s}_{n,n+1}^z} =(\beta-1)/(2\beta+2)$, serving as an effective unidirectional electric field that generates the accumulation of dipole charge for particles, i.e., $\operatorname{Tr}(\rho_{\text{ss}, N} D)\ne0$. 

Furthermore, $ \rho_{\text{ss}}$ does not depend on the interaction strength $J$. Instead, the local gauge symmetry of $ H$ plays an indispensable role in constructing the double-space transformation $\mathcal{T}$. This result indicates that $ \rho_{\text{ss}}$ is robust against the symmetry-preserving disorders in $ H$, such as random potentials $\delta H=\sum_{n=1}^Lh_n\tau_n^z+\sum_{n=1}^{L-1}h^\prime_{n,n+1} s_{n,n+1}^z$ or long-range interactions  $\delta H^\prime=\sum_{n=1}^{L-2}J^\prime_n( \tau_n^+ s_{n,n+1}^+  s_{n+1,n+2}^+\tau_{n+2}^{-}+\text{h.c.})$. This symmetry-protected feature distinguishes the exact steady-state skin mode from the single-particle NHSE in noninteracting systems, where strong disorder typically destroys skin modes and instead leads to Anderson localization. 

The exact construction leading to Eq. \eqref{eq:exact_steady} offers a general approach to obtaining eigenoperators of a large class of Liouvillian superoperators. With the Liouvillian $\mathcal{L}=\mathcal{L}_H+\mathcal{L}_D$, if we find a double-space similarity transformation $\mathcal{T}_\lambda$ such that 
\begin{equation}
\mathcal{T}_\lambda\mathcal{L}_H\mathcal{T}_\lambda^{-1}=\mathcal{L}_H,\ \mathcal{T}_\lambda\mathcal{L}_D\mathcal{T}_\lambda^{-1}[ I]=\lambda  I,
\end{equation}
we can obtain an eigenoperator $\rho=\mathcal{T}_\lambda^{-1}[ I]$ of $\mathcal{L}$ with the eigenvalue $\lambda$. In the supplemental material \cite{supp_LGT}, we show that this approach systematically provides exponentially many eigenoperators of $\mathcal{L}$ in dissipative QLM, with eigenvalues equally spaced on the real axis [black points in Fig.\ref{fig:spectrum}(b)].  We remark that the equally spaced eigenvalues are reminiscent of exact many-bo
dy scar states in a large class of Hermitian systems \cite{Ramirez2015Entanglement,Langlett2022Rainbow,Wildeboer2022scar}. We also apply this method to other types of jump operators in dissipative QLM \cite{supp_LGT}. These demonstrations further indicate the general applicability of our exact method.

\begin{figure}
    \centering
    \includegraphics{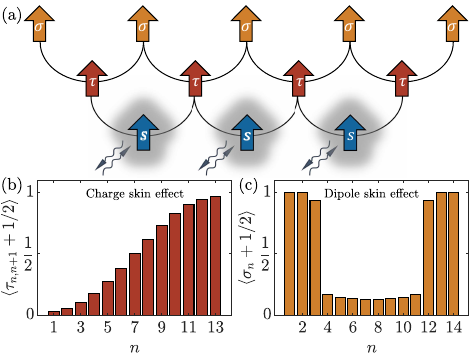}
    \caption{(a) Generalized dissipative QLM. (b,c) Steady-state distribution of the middle and top layers. With $L=14$ and $\beta=3$, we show the results in the charge sector $( N_{H^\prime}, D_{H^\prime})=(0,0)$. }
    \label{fig:hierarchical}
\end{figure}
\emph{Hierarchical skin effect.--} Fig. \ref{fig:spectrum} elucidates that the biased dissipation of link spins causes the many-body skin effect of site spins and increases their dipole moments. Interestingly, this setup provides a general framework to generate a new type of many-body NHSE, called the hierarchical skin effect, where the steady state will accumulate nonzero multipole moments for different subsystems.

We consider a generalized dissipative QLM [Fig. \ref{fig:hierarchical}(a)], whose Hamiltonian is defined for three species of spin-$\frac{1}{2}$ operators:
\begin{equation}\label{eq:generalized_Ham}
  H^\prime=J_1\sum_{n=1}^{L-1}({\sigma}_n^+ {\tau}_{n,n+1}^+{\sigma}_{n+1}^-+\text{h.c.})+J_2\sum_{n=1}^{L-2}({\tau}_{n,n+1}^+ {s}_{n+1}^+{\tau}_{n+1,n+2}^-+\text{h.c.}).
\end{equation}
Here, $L$ is the number of $\sigma$ spins on the top layer. Likewise, the middle layer has $L-1$ $\tau$ spins and the bottom layer consists of $L-2$ $s$ spins. We denote $ H_1\equiv  H^\prime|_{J_2=0}$ and $ H_2\equiv  H^\prime|_{J_1=0}$, which are quantum-link Hamiltonians between neighboring layers.   Like Eq.\eqref{eq:hamiltonian}, $ H_1$ has a local U(1) gauge symmetry generated by $ G_{n}^{\sigma\tau}= \sigma_n^z- \tau_{n,n+1}^z+ \tau_{n-1,n}^z$ for $1<n<L$, with boundary corrections $ G_{1}^{\sigma\tau}= \sigma_1^z- \tau_{1,2}^z$ and $ G_{L}^{\sigma\tau}= \sigma_L^z+ \tau_{L-1,L}^z$. Similarly, the local U(1) gauge symmetry generators for $ H_2$ are given by  $ G_{n,n+1}^{\tau s}= \tau_{n,n+1}^z- s_{n+1}^z+ s_{n}^z$ for $1<n<L-1$, $ G_{1,2}^{\tau s}= \tau_{1,2}^z- s_{2}^z$, and $ G_{L-1,L}^{\tau s}= \tau_{L-1,L}^z+ s_{L-1}^z$. However, these symmetry generators for $ H_1$ or $ H_2$  are no longer conserved when considering the total Hamiltonian $ H^\prime= H_1+ H_2$. Instead, the local symmetry generators in $ H^\prime$ involve spin operators in all layers, which are given by $ G_n^{H^\prime}= \sigma_n^z- G^{\tau s}_{n,n+1}+ G^{\tau s}_{n-1,n}$ for $1<n<L$, $ G_{1}^{H^\prime}= \sigma_1^z- G^{\tau s}_{1,2}$, and $ G_{L}^{H^\prime}= \sigma_L^z+ G^{\tau s}_{L-1,L}$. It is easy to show that $[ G_n^{H^\prime}, H^\prime]=0$. Consequently,  $ G_n^{H^\prime}$ plays a similar role to the local U(1) gauge symmetry generators in the original quantum-link Hamiltonian.

With $\beta=\gamma_u/\gamma_d$, we added biased jump operators $ L_{n}^{(u)}=\sqrt{\gamma_u}{s}_{n}^{+}$ and $ L_{n}^{(d)}=\sqrt{\gamma_d} {s}_{n}^{-}$ to $s$ spins [Fig.\ref{fig:hierarchical}(a)], where $n=2,3,\cdots,L-1$. The dynamics of this system is generated by the Lindblad master equation in Eq.\eqref{eq:lindblad}.  For convenience, we express the Liouvillian superoperator as $\mathcal{L}=\mathcal{L}_{H^\prime}+\mathcal{L}_{D}$, with $\mathcal{L}_{H^\prime}$ given by the Hamiltonian $ H^\prime$ and $\mathcal{L}_{D}$ corresponding to the dissipations of $s$ spins, respectively. 

The way to obtain the steady state is to construct a double-space similarity transformation $\mathcal{T}$, defined as $\mathcal{T}[\rho]=( T_{\sigma}\otimes  T_{\tau}\otimes  T_s)\rho$, such that $\mathcal{T}\mathcal{L}\mathcal{T}^{-1}=\mathcal{L}_{H^\prime}+\mathcal{L}_{D}^\dagger$ . When $ H_1=0$, the top layer is decoupled. The rest of the system is the same as Fig.\ref{fig:spectrum}(a). The results above Eq.\eqref{eq:exact_steady} shows that $ T_\tau\otimes  T_s=\exp[-\sum_{n=1}^{L-1}(\ln\alpha+n\ln\beta) G_{n,n+1}^{\tau s}]$ with $\alpha$ being a free parameter. A nonzero $ H_1$ constrains the form of $ T_\sigma$, indicating that $ T_\sigma\otimes T_\tau$ should keep $ {H}_1$ invariant and be compatible with the above $ T_\tau\otimes  T_s$. As a result, $ T_\sigma\otimes T_\tau=\exp\{-\sum_{n=1}^L[\ln\alpha'+\sum_{m=0}^{n-1}(\ln\alpha+m\ln\beta)]  G^{\sigma\tau}_n\}$ with $\alpha'$ being another free parameter. Combining these results yields the steady state
\begin{equation}\label{eq:generalized_steady_state}
    \rho_{\text{ss}}\propto\exp\left[\sum_{n=1}^L\left(\ln\alpha'+n\ln\alpha+\frac{n(n-1)}{2}\ln\beta\right)  G^{H^\prime}_n\right].
\end{equation}

The free parameters $\alpha$ and $\alpha^\prime$ reveal two strong global symmetries that do not involve $s$-spin operators and thus commute with $ L_n^{(u/d)}$. The first one, $ N_{H^\prime}=\sum_{n=1}^L G^{H^\prime}_n=\sum_{n=1}^L\sigma_n^z$, is the total $z$-component of $\sigma$ spins. The second one, $ D_{H^\prime}=\sum_{n=1}^Ln G^{H^\prime}_n=\sum_{n=1}^Ln\sigma_n^z+\sum_{n=1}^{L-1} \tau_{n,n+1}^z$, consists of the dipole moment of $\sigma$ spins and the total $z$-component of $\tau$ spins.  After projection into each symmetry sector, we obtain the exact symmetry-resolved steady states that exclusively depend on the dissipation parameter $\beta$. A typical steady-state distribution is shown in Fig.\ref{fig:hierarchical}. Intuitively, the polarization of $s$ spins causes the many-body skin effect for $\tau$ spins [Fig.\ref{fig:hierarchical}(b)], generating a nonzero dipole moment in the middle layer. Subsequently, the asymmetric distribution of $\tau$ spins effectively serves as a spatially dependent electric field for $\sigma$ spins. The opposite polarizations of $\tau$ spins [Fig.\ref{fig:hierarchical}(b)] transport the polarizations of $\sigma$ spins in different directions, making the $z$-component of $\sigma$ spins accumulate at two boundaries simultaneously  [Fig.\ref{fig:hierarchical}(c)].  As a result, a nonzero quadrupole moment for $\sigma$ spins reveals the existence of dipole skin effect in the top layer \cite{gliozzi2024manybody}.

More layers with interlayer quantum-link interactions can be included in the open quantum system, such that a hierarchy of many-body NHSE can be established in the exact steady state, characterized by nonzero multipole moments for different layers. This intriguing feature distinguishes the generalized dissipative QLM from the previous result on the many-body NHSE for multipoles in non-Hermitian Hamiltonians \cite{gliozzi2024manybody}. Whereas the latter necessitates a particular $m$-pole conserving non-Hermitian interaction to produce the $(m+ 1)$-th multipole moment, our framework for hierarchical skin effect requires merely 3-local interactions and on-site dissipations that are more feasible to realize experimentally, simultaneously creating multipole skin effects of any order in different subsystems.

\emph{Possible experiments.--}Our model offers a practical method in engineering nonreciprocal hoppings in a quantum many-body system. One possible platform is the extended Bose-Hubbard simulator in cold-atom systems~\cite{yang2020observation}. By introducing a spatially modulated potential and specifying the model parameters, an effective bosonic Hamiltonian can be engineered in the form \( H_{\text{eff}} = J \sum_{n} ( b_{2n-1} (b_{2n}^\dagger)^2 b_{2n+1} + \text{h.c.} ) \), with the dynamics restricted to the Hilbert subspace with zero or single (double) boson occupations on odd (even) sites. Within this subspace, the model describes a U(1) lattice gauge theory, where the odd (even) sites represent matter (gauge) fields. The detailed derivation and its mapping to the quantum link model [Eq.\eqref{eq:hamiltonian}]  can be found in Ref.~\cite{yang2020observation}. Under the same mapping, the spin-flipping jump operators in Eq. \eqref{eq:biased_jump} can be translated into bosonic gain/loss jump operators   \cite{Syassen_2008,Tomita2019Dissipative,sieberer2016keldysh} on even sites, although the dissipative dynamics need to be restricted in the above-mentioned subspace. Another promising platform is the superconducting circuit, where a native three-body quantum-link interaction [Eq.\eqref{eq:hamiltonian}] has recently been realized \cite{busnaina2025native} and there exist known strategies for introducing gain/loss jump operators into superconducting qubits \cite{Blais2021circuit}. 

\emph{Discussions.--} In conclusion, we have constructed gauge-theoretical models with exactly solvable steady states exhibiting NHSE.  Our results shed light on novel nonequilibrium phases tied to gauge symmetries. For instance, the approach to induce the NHSE in dissipative many-body systems and develop the hierarchical skin effect for multipoles can be easily applied to U(1) QLMs in any dimensions with higher spins \cite{Felser2020two, Cardarelli2020deconfining, VanDamme2022Dynamical, Zache2022toward, Desaules2023Prominent, Desaules2023weak}. An example of  2D NHSE in dissipative QLMs is shown in the supplemental material \cite{supp_LGT}. Another interesting future direction is to extend the dissipative lattice gauge theory to non-Abelian cases \cite{Banerjee2013UN,Mezzacapo2015nonabelian}. It is also intriguing to investigate the dissipative gauge fields using Keldysh field theory \cite{sieberer2016keldysh,sieberer2023universality}  and topological field theory \cite{Kawabata2021topological}.

\emph{Acknowledgement.--} We thank He-Ran Wang, Fei Song, Shunyu Yao, Zhou-Quan Wan, and Hongzheng Zhao for their helpful discussions. This work is supported by the National Natural Science Foundation of China (Grants No. 12125405), the National Key R\&D Program of China (No. 2023YFA1406702), and Quantum Science and Technology-National Science and Technology Major Project (Grant No. 2021ZD0302502). Biao Lian is supported by the National Science Foundation through Princeton University’s Materials Research Science and Engineering Center DMR-2011750, and the National Science Foundation under award DMR-2141966. Additional support is provided by the Gordon and Betty Moore Foundation through Grant GBMF8685 towards the Princeton theory program.

\bibliography{ref_LGT}

\end{document}

\title{Supplemental Material:  Many-body non-Hermitian skin effect with exact steady states in the dissipative quantum link model}	
	
\author{Yu-Min Hu}
\affiliation{Institute for Advanced Study, Tsinghua University, Beijing,  100084, China}
\affiliation{Max Planck Institute for the Physics of Complex Systems, N\"{o}thnitzer Str. 38, 01187 Dresden, Germany}
\author{Zijian Wang}
\affiliation{Institute for Advanced Study, Tsinghua University, Beijing,  100084, China}

\author{Biao Lian}
\altaffiliation{biao@princeton.edu}

\affiliation{Department of Physics, Princeton University, Princeton, New Jersey 08544, USA}

\author{Zhong Wang}
\altaffiliation{wangzhongemail@tsinghua.edu.cn}

\affiliation{Institute for Advanced Study, Tsinghua University, Beijing,  100084, China}  

%\date{\today}

\maketitle

\section{Relaxation dynamics in the dissipative QLM}
\begin{figure}[t]
   \centering
    \includegraphics[width=8.5cm]{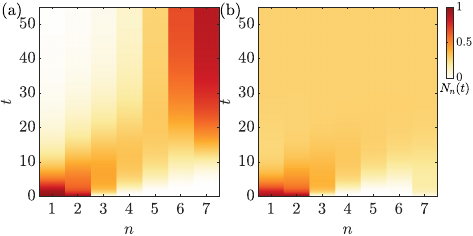}
    \caption{The quench dynamics from the initial pure state with the filling configuration  $\ket{\bullet\bullet\circ\circ\circ\circ\circ}$  on lattice sites and all downward spins $\ket{\downarrow}$ on lattice links. Here, $ \tau_n^z\ket{\bullet}=1/2\ket{\bullet}$ and $ \tau_n^z\ket{\circ}=-1/2\ket{\circ}$. The colormap shows $N_n(t)=\operatorname{Tr}[\rho(t)(\tau^z_n+1/2)]$. (a) takes the open boundary conditions, while (b) takes the periodic boundary conditions. Parameters: $J=1,\ \gamma_u=3,\ \gamma_d=1,\ N=2,\ L=7$.}
    \label{fig:dynamics}
\end{figure}

In this section, we show the quench dynamics of the open quantum system in Fig.1(a) of the main text. Starting from a specific initial state,  this open quantum system evolves into different steady states determined by boundary conditions. The dissipative gauge fields effectively induce a chiral motion of particles on lattice sites. Under periodic boundary conditions (PBC), the chiral motion of particles is visible in the short-time dynamics, yet it quickly evolves into a uniform steady state that preserves translation symmetry [Fig. \ref{fig:dynamics}(b)]. In contrast, the chiral motion under open boundary conditions (OBC) renders particles accumulated at the boundary, resulting in many-body NHSE in the steady state [Fig.\ref{fig:dynamics}(a)]. These dynamical features provide a prominent signal to detect many-body non-Hermitian skin effect in the dissipative lattice gauge theory.

\section{Point-gap topology of many-body Liouvillian}
It is known that, in many-body non-Hermitian systems, there exists a correspondence between the nontrivial point-gap topology for many-body PBC spectrums and the existence of many-body NHSE under OBC \cite{Kawabata2022manybody}. In a finite PBC system, the non-Hermitian many-body spectrum with nontrivial point-gap topology exhibits a spectrum winding with the tuning of a twisted boundary condition \cite{Kawabata2022manybody}. In this section, we show that such a topological correspondence is also applicable to predict the many-body NHSE in open quantum many-body systems.  We show that there are two types of twisted boundaries in open quantum systems, and only one of them induces a nontrivial spectrum winding pattern for the Liouvillian considered in our work.

The first type involves imposing a twisted boundary on the Hamiltonian while maintaining the Liouvillian superoperator structure. Nevertheless, such a twisted boundary will not change the Liouvillian spectrum of our model.  To see this, we consider periodic boundary conditions and modify the quantum-link Hamiltonian  $H_{\text{PBC}}=\sum_{n=1}^L(J{\tau}_n^+ {s}_{n,n+1}^+{\tau}_{n+1}^-+\text{h.c.})$ to 
\begin{equation}
   H_{\text{PBC}}(\phi)=   H_{\text{OBC}}+Je^{i\phi}{\tau}_L^+ {s}_{L,1}^+{\tau}_{1}^-+Je^{-i\phi}{\tau}_L^- {s}_{L,1}^-{\tau}_{1}^+, 
\end{equation} where $H_{\text{OBC}}=\sum_{n=1}^{L-1}J({\tau}_n^+ {s}_{n,n+1}^+{\tau}_{n+1}^-+\text{h.c.})$ and $\phi$ corresponds to the twisted boundary condition between the first and last sites. For later convenience, we define $H_{\text{twist}}({\phi})=Je^{i\phi}{\tau}_L^+ {s}_{L,1}^+{\tau}_{1}^-+Je^{-i\phi}{\tau}_L^- {s}_{L,1}^-{\tau}_{1}^+$. Under a unitary transformation $Us_{L,1}^{\pm} U^\dagger= e^{\mp i\phi}s_{L,1}^{\pm}$ with $U=\exp(-i{\phi}s_{L,1}^z)$ acting on the Hilbert space of the boundary link, we find that $UH_{\text{PBC}}(\phi)U^\dagger = H_{\text{PBC}}(0)$, which means that $H_{\text{PBC}}(\phi)$ and $H_{\text{PBC}}(0)$ share the same spectrum yet have distinct eigenstates.  Meanwhile, the unitary transformation will also transfer the density matrix $\rho$ into $U\rho U^\dagger$.

With the twisted boundary condition in the Hilbert space, the full Liouvillian superoperator now depends on $\phi$:
\begin{equation}
\begin{split}
    \mathcal{L}_\phi[\rho]=&-i[H_{\text{PBC}}(\phi),\rho]+\sum_\mu 2L_\mu\rho L_\mu^\dagger-\{L_\mu^\dagger L_\mu,\rho\}\\
    =&\mathcal{L}_{\text{OBC}}[\rho]-i[H_{\text{twist}}({\phi}),\rho]\\
    &+\sum_{i=u,d} 2L_{L,1}^{(i)}\rho (L_{L,1}^{(i)})^\dagger-\{(L_{L,1}^{(i)})^\dagger L_{L,1}^{(i)},\rho\}.
\end{split}
\end{equation} In the above equation, we recall that the jump operators are given by $L_{n,n+1}^{(u)}=\sqrt{\gamma_u}s_{n,n+1}^+$ and $L_{n,n+1}^{(d)}=\sqrt{\gamma_d}s_{n,n+1}^-$. $\mathcal{L}_{\text{OBC}}$ is the Liouvillian under open boundary conditions.  Under the aforementioned unitary transformation $U$ in the Hilbert space, the Hamiltonian $H_{\text{PBC}}(\phi)$ becomes $H_{\text{PBC}}(0)$  and  the jump operators become $L_{L,1}^{u}=\sqrt{\gamma_u}e^{-i\phi}s_{L,1}^+$ and $L_{L,1}^{d}=\sqrt{\gamma_d}e^{+i\phi}s_{L,1}^-$.  However, a phase rotation in jump operators does not affect the Liouvillian spectrum, since the jump operators in the Lindblad master equation always appear in Hermitian conjugate pairs and the phase factor cancels out. As a result, $\mathcal{L}_\phi$ becomes $\mathcal{L}_0$ under a unitary transformation, and therefore, shares the same Liouvillian spectrum as $\mathcal{L}_0$. This means that inserting a twisted boundary condition at the Hamiltonian level will not affect the Liouvillian spectrum.

We can also express $\mathcal{L}_\phi$ in the double Hilbert space:
\begin{equation}
    \begin{split}
\mathcal{L}_\phi=&\mathcal{L}_{\text{OBC}}-iH_{\text{twist}}(\phi)\otimes I+iI\otimes [H_{\text{twist}}(\phi)]^T\\
&+2\gamma_u s_{L,1}^{+}\otimes s_{L,1}^{+}-s_{L,1}^{-} s_{L,1}^{+}\otimes I - I\otimes s_{L,1}^{-} s_{L,1}^{+}\\
&+2\gamma_d s_{L,1}^{-}\otimes s_{L,1}^{-}-s_{L,1}^{+} s_{L,1}^{-}\otimes I - I\otimes s_{L,1}^{+} s_{L,1}^{-}. 
    \end{split}
\end{equation} In the double Hilbert space, the aforementioned unitary transformation $U$ becomes a superoperator $\mathcal{U}=U\otimes U^*=\exp(-i{\phi}s_{L,1}^z)\otimes \exp(+i{\phi}s_{L,1}^z)$, which leads to $\mathcal{U}\mathcal{L}_\phi\mathcal{U}^\dagger=\mathcal{L}_0$.

\begin{figure}[t]
    \centering
    \includegraphics[width=8.5cm]{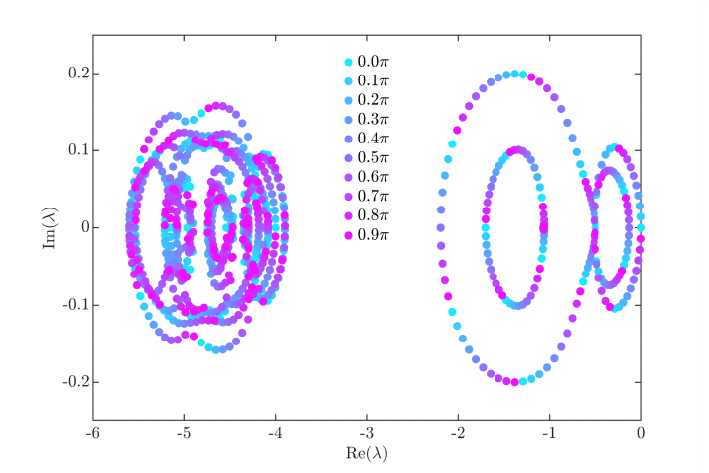}
    \caption{The periodic spectrum of $\mathcal{L}_\phi^\prime$ with a nonzero $\phi$. These spectra are numerically obtained by the exact diagonalization method in the PBC sector $\mathcal{G}_n=0$ and $N=2$ with parameters $L=7$, $J=1$, $\gamma_u=2.4$, and $\gamma_d=1.6$. For a better illustration of spectrum winding, we only show a part of the Liouvillian spectrum. The full PBC spectrum with $\phi=0$ is presented in Fig. 1(b) of the main text.}
    \label{sfig:winding}
\end{figure}

The second way is to directly view the Liouvillian superoperator $\mathcal{L}$ as a non-Hermitian matrix and treat the ``ket" and ``bra" sides at the same (rather than opposite in the previous case) level. Therefore, we impose a twisted boundary condition in the double Hilbert space for the PBC Liouvillian superoperator, which corresponds to the symmetry flux of strong global $U(1)$ symmetry.  With this picture in mind, we consider the following generalized superoperator in the double Hilbert space:
\begin{equation}
    \begin{split}
\mathcal{L}_\phi^\prime=&\mathcal{L}_{\text{OBC}}-iH_{\text{twist}}(\phi)\otimes I+iI\otimes H_{\text{twist}}(\phi)\\
&+2\gamma_u s_{L,1}^{+}\otimes s_{L,1}^{+}-s_{L,1}^{-} s_{L,1}^{+}\otimes I - I\otimes s_{L,1}^{-} s_{L,1}^{+}\\
&+2 \gamma_d s_{L,1}^{-}\otimes s_{L,1}^{-}-s_{L,1}^{+} s_{L,1}^{-}\otimes I - I\otimes s_{L,1}^{+} s_{L,1}^{-}. 
    \end{split}
\end{equation} We remark that $H_{\text{twist}}(0)=[H_{\text{twist}}(0)]^T$ but $ H_{\text{twist}}(\phi)\ne[H_{\text{twist}}(\phi)]^T=H_{\text{twist}}(-\phi)$. Namely, the matrix transport operation and the insertion of a twisted boundary do not commute with each other. In this sense, $\mathcal{L}_\phi^\prime$ is not equivalent to the above $\mathcal{L}_\phi$ and does not represent a Lindblad master equation anymore.

The resulting PBC spectrum of $\mathcal{L}_\phi^\prime$ for different $\phi$ is shown in Fig. \ref{sfig:winding}. With the increase of $\phi$, the PBC spectrum will periodically move in the complex plane and encircle a nonzero area. As shown in Fig. \ref{sfig:winding}, the periodicity of recovering the $\mathcal{L}_0$ spectrum is given by $\pi$. This can be understood as follows. We can perform a double-space unitary transformation $\mathcal{U}^\prime=U\otimes U^T=\exp(-i{\phi}s_{L,1}^z)\otimes \exp(-i{\phi}s_{L,1}^z)$, which differs from the above $\mathcal{U}$.  The double-space unitary transformation $\mathcal{U}^\prime$ leads to  $e^{\pm i\phi}s_{L,1}^\pm\otimes I\to s_{L,1}^\pm\otimes I $ and $e^{\pm i\phi}I\otimes s_{L,1}^\pm\to I\otimes s_{L,1}^\pm$. As a result, we obtain
\begin{equation}
    \begin{split}
        \mathcal{L}_\phi^{\prime\prime}\equiv& \mathcal{U}^\prime\mathcal{L}_\phi^\prime (\mathcal{U}^\prime)^\dagger=\mathcal{L}_{\text{OBC}}-iH_{\text{twist}}(0)\otimes I+iI\otimes H_{\text{twist}}(0)\\
&+2\gamma_u e^{-2i\phi}s_{L,1}^{+}\otimes s_{L,1}^{+}-s_{L,1}^{-} s_{L,1}^{+}\otimes I - I\otimes s_{L,1}^{-} s_{L,1}^{+}\\
&+2\gamma_d e^{+2i\phi}s_{L,1}^{-}\otimes s_{L,1}^{-}-s_{L,1}^{+} s_{L,1}^{-}\otimes I - I\otimes s_{L,1}^{+} s_{L,1}^{-}.      
    \end{split}
\end{equation}
which has a $\pi$-periodicity. 

In conclusion, the nontrivial spectral winding of the PBC Liouvillian superoperator $\mathcal{L}_\phi^\prime$ signifies the topological origin of many-body skin modes under OBC.
\section{Exact eigenoperators in the dissipative QLM}\label{sec:exact} 
The exact construction leading to Eq. (5) in the main text offers a general approach to obtaining eigenoperators of a large class of Liouvillian superoperators in the form $\mathcal{L}=\mathcal{L}_H+\mathcal{L}_D$. The key is to find a double-space similarity transformation $\mathcal{T}$ such that $\mathcal{T}\mathcal{L}_H\mathcal{T}^{-1}=\mathcal{L}_H$ and $\mathcal{T}\mathcal{L}_D\mathcal{T}^{-1}[ I]=\lambda  I$. Then we can obtain an eigenoperator $\rho=\mathcal{T}^{-1}[ I]$ of $\mathcal{L}$ with the eigenvalue $\lambda$. In this section, we show how to systematically construct exponentially many eigenoperators of $\mathcal{L}=\mathcal{L}_H+\mathcal{L}_D$ in the dissipative QLM shown in Fig.1(a) of the main text, with $\mathcal{L}_H$ determined by the quantum link Hamiltonian in Eq. (1) of the main text and $\mathcal{L}_D$ given by Eq. (3) of the main text.

We start from $\mathcal{L}_H=0$. Instead of steady states, we discuss the dissipative eigenstates $ \varrho$ of $\mathcal{L}_D$ with  $ \varrho=\varrho_\tau\otimes\varrho_s$. Here $\varrho_\tau$ is an arbitrary operator in the site Hilbert space and $\varrho_s=\otimes_{n=1}^{L-1}\varrho_{n,n+1}$ defined on decoupled links. For each decoupled link spin, there are four choices of eigenoperators $\varrho_{n,n+1}$ satisfying $\mathcal{L}_{D,n,n+1}[\varrho_{n,n+1}]=\lambda\varrho_{n,n+1}$, where $\mathcal{L}_{D,n,n+1}$ consists only of jump operators $ L_{n,n+1}^{(u)}=\sqrt{\gamma_u}{s}_{n,n+1}^{+} $ and $ L_{n,n+1}^{(d)}=\sqrt{\gamma_d} {s}_{n,n+1}^{-}$ on the $(n,n+1)$-link. We hereafter focus on two diagonal eigenoperators: $ \varrho_{n,n+1}^{(0)}=\exp(\ln\beta_0 s_{n,n+1}^z)$ with $\beta_0=\gamma_u/\gamma_d$ and $\lambda_0=0$; $ \varrho_{n,n+1}^{(1)}=\exp(\ln\beta_1 s_{n,n+1}^z)$ with $\beta_1=-1$ and $\lambda_1=-2(\gamma_u+\gamma_d)$. With $\varrho_{n,n+1}$ chosen in $\{  \varrho_{n,n+1}^{(0)}, \varrho_{n,n+1}^{(1)}\}$, the link part $\varrho_s$ of a possible eigenoperator of $\mathcal{L}_D$ can take the form $\varrho_{s,{\mathbf{k}}}=\otimes_{n=1}^{L-1} \varrho_{n,n+1}^{(k_n)}=\exp(\sum_{n=1}^{L-1}\ln\beta_{k_n} s_{n,n+1}^z)$, where ${\mathbf{k}}$ denotes a bit string $k_1k_2\cdots k_{L-1}$ with $k_n\in\{0,1\}$. Namely. $\mathcal{L}_D[\varrho_\tau\otimes\varrho_{s,\mathbf{k}}]=\lambda_{\mathbf{k}}\varrho_\tau\otimes\varrho_{s,\mathbf{k}}$ and the eigenvalue $\lambda_{\mathbf{k}}$ is given by $\lambda_{\mathbf{k}}=\sum_{n=1}^{L-1}\lambda_{k_n}=-2(\gamma_d+\gamma_u)K$ with $K=\sum_{n=1}^{L-1}k_n$. Apparently, the eigenvalue determined by $K$ has a $C_{L-1}^K$-fold degeneracy.

 We turn to the case with $\mathcal{L}_H\ne0$. As discussed in the main text, we shall find the double-space transformation $\mathcal{T}_{\mathbf{k}}$ satisfying $\mathcal{T}_{\mathbf{k}}\mathcal{L}_H\mathcal{T}_{\mathbf{k}}^{-1}=\mathcal{L}_H$ and $\mathcal{T}_{\mathbf{k}}[\varrho_\tau\otimes\varrho_{s,\mathbf{k}}]= I$. The compatible transformation is found to be $\mathcal{T}_{\mathbf{k}}[\varrho]= T_{\mathbf{k}}\varrho$ with $ T_{\mathbf{k}}=\exp[-\sum_{n=1}^{L}(\ln\alpha+\sum_{i=1}^{n-1}\ln\beta_{k_n}) G_n]$. Here, $ G_n$ are local gauge symmetry generators of the quantum-link Hamiltonian. As a result, the exact eigenoperator of $\mathcal{L}$ is given by
 \begin{equation}\label{eq:exact_operator}
     \varrho_{\mathbf{k}}=\exp\left[\sum_{n=1}^{L}\left(\ln\alpha+\sum_{i=1}^{n-1}\ln\beta_{k_i}\right) G_n\right]. 
 \end{equation}
The corresponding eigenvalue is provided by the above $\lambda_{\mathbf{k}}=-2(\gamma_d+\gamma_u)K$ with $K=\sum_{n=1}^{L-1}k_n$. Projecting $\varrho_{\mathbf{k}}$ into the $N$-particle Hilbert space, we obtain eigenoperators independent of the free parameter $\alpha$. After resolving the strong global U(1) symmetry, these exact eigenoperators have a degeneracy of $(L+1)C_{L-1}^K$. 

Remarkably, even though it is a difficult task to obtain the full many-body spectrum of $\mathcal{L}$, we can employ the gauge structure of the quantum-link Hamiltonian to exactly construct $2^{L-1}$ eigenoperators in dissipative QLM. These exponentially many eigenoperators are also independent of the interaction strength $J$, relying only on the local gauge generators of $ H$. Their eigenvalues, shown by black points in Fig. 1(a) of the main text, are equally spaced on the real axis.

\section{Other types of jump operators}
In this section, we discuss other choices of jump operators in the dissipative quantum link model. In all cases, we take open boundary conditions. We will show how the different types of jump operators affect the steady states.

\subsection*{$s_{n,n+1}^x$-like jump operators}
We first study the $s_{n,n+1}^x$-like jump operators: 
\begin{equation}\label{seq:xjump}
    L_{n,n+1}=\sqrt{\gamma_u}s_{n,n+1}^+ + \sqrt{\gamma_d}s_{n,n+1}^{-}.
\end{equation}
These jump operators become $L_{n,n+1}\propto s_{n,n+1}^x$ when $\gamma_d=\gamma_u$, which is the reason for the name ``$s_{n,n+1}^x$-like". Here, we consider a single jump operator acting on each lattice link, which is different from the two biased jump operators introduced in Eq. (3) of the main text. This single jump operator also breaks the strong U(1) gauge symmetry. Moreover, the eigenstate construction method in Sec. \ref{sec:exact} still applies to this open quantum system  $\mathcal{L}=\mathcal{L}_H+\mathcal{L}_D$ , with $\mathcal{L}_H$ determined by the quantum link Hamiltonian in Eq. (1) of the main text and $\mathcal{L}_D$ determined by jump operators in Eq. \eqref{seq:xjump}. 

We first consider $\mathcal{L}_H=0$. We discuss the eigenstates $ \varrho$ of $\mathcal{L}_D$ in the form  $ \varrho=\varrho_\tau\otimes\varrho_s$. $\varrho_s=\otimes_{n=1}^{L-1}\varrho_{n,n+1}$ is defined on decoupled links. For each decoupled link spin, there are four choices of eigenoperators $\varrho_{n,n+1}$ satisfying $\mathcal{L}_{D,n,n+1}[\varrho_{n,n+1}]=\lambda\varrho_{n,n+1}$, where $\mathcal{L}_{D,n,n+1}$ is generated by $ L_{n,n+1}=\sqrt{\gamma_u}{s}_{n,n+1}^{+} +\sqrt{\gamma_d} {s}_{n,n+1}^{-}$ on the $(n,n+1)$-link.  To proceed, we focus on two diagonal eigenoperators that commute with  $ G_n={\tau}_n^z-( s_{n,n+1}^z- s_{n-1,n}^z)$: $ \varrho_{n,n+1}^{(0)}=\exp(\ln\beta_0 s_{n,n+1}^z)$ with $\beta_0=\gamma_u/\gamma_d$ and $\lambda_0=0$; $ \varrho_{n,n+1}^{(1)}=\exp(\ln\beta_1 s_{n,n+1}^z)$ with $\beta_1=-1$ and $\lambda_1=-2(\gamma_u+\gamma_d)$.

Remarkably, these two diagonal eigenoperators for decoupled link spins are the same as those of two biased spin-flipping jump operators in Sec. \ref{sec:exact}.  We thus follow the same procedure here. Including a nonzero Hamiltonian $\mathcal{L}_H\ne0$ and constructing the same double-space transformation as Sec. \ref{sec:exact}, we get the exact eigenoperator of $\mathcal{L}$:
 \begin{equation}\label{eq:exact_operator}
     \varrho_{\mathbf{k}}=\exp\left[\sum_{n=1}^{L}\left(\ln\alpha+\sum_{i=1}^{n-1}\ln\beta_{k_i}\right) G_n\right],
 \end{equation}
where ${\mathbf{k}}$ denotes a bit string $k_1k_2\cdots k_{L-1}$ with $k_n\in\{0,1\}$.  The corresponding eigenvalue is provided by the above $\lambda_{\mathbf{k}}=-2(\gamma_d+\gamma_u)K$ with $K=\sum_{n=1}^{L-1}k_n$.  After resolving the strong global U(1) symmetry, these exact eigenoperators have a degeneracy of $(L+1)C_{L-1}^K$. 

Specifically, when $k_1=k_2=\cdots=k_{L-1}=0$, we obtain the steady state in this case:
 \begin{equation}
     \rho_{\text{ss}}=\exp\left[\sum_{n=1}^{L}\left(\ln\alpha+(n-1)\ln\beta_{0}\right) G_n\right]. 
 \end{equation}
 After rescaling the free parameter $\alpha\to\alpha\beta_0$, we obtain the same steady state as Eq. (5) of the main text, which exhibits many-body non-Hermitian skin effect when $\beta_0=\gamma_u/\gamma_d\ne1$. 
 
However, when $\beta_0=\gamma_u/\gamma_d=1$, the jump operator in Eq. \eqref{seq:xjump} becomes $L_{n,n+1}\propto s_{n,n+1}^x$, which is symmetric in two spin-flipping directions. Because this jump operator is a symmetric Hermitian operator, the steady state becomes the identity matrix in each symmetry sector labeled by the total particle number $N=\sum_{n=1}^L(\tau_n^z+1/2)$. In this case, the non-Hermitian skin effect is absent.

As a final remark, we point out that although the $s_{n,n+1}^x$-like jump operators share the same set of exact eigenstates as the two biased spin-flipping jump operators considered in Eq. (3) of the main text,  these exactly solvable eigenstates represent only a small portion of the full Liouvillian spectrum.  The majority of eigenstates differ between the two cases, and these differences can lead to distinct impacts on the relaxation dynamics, even though both systems evolve toward the same steady state.

\subsection*{Dephasing}
The second jump operator considered here is the dephasing operator:
\begin{equation}\label{seq:dephasing}
    L_{n,n+1}=\sqrt{\gamma}s_{n,n+1}^{z}.
\end{equation}
Since this is a Hermitian jump operator, the steady state is the identity operator in each symmetry sector. Notably, since $[s_{n,n+1}^z, G_m]=0$, this jump operator preserves the strong U(1) local symmetry. In this sense, there exists one steady state--specifically, the identity operator--in each gauge sector specified by the eigenvalues of $G_1,\ G_2,\cdots, G_L$.   Such an identity operator is not expected to exhibit many-body non-Hermitian skin effect. Unlike the case in the main text, where there are $O(N)$ steady states specified by the total particle number, here the number of steady states specified by strong U(1) gauge symmetry grows exponentially with the system size. 

\section{2D dissipative quantum link model}
In this section, we discuss the many-body non-Hermitian skin effect in the two-dimensional (2D) dissipative quantum link model.  We consider a 2D square lattice of size \( L \times L \). Lattice sites are labeled by integer coordinates \( (x, y) \), while the links in the vertical and horizontal directions are denoted by \( (x+\frac{1}{2}, y) \) and \( (x, y+\frac{1}{2}) \), respectively, corresponding to the upper and right links of each site. The 2D quantum link Hamiltonian is
\begin{equation}
    H=\sum_{x,y=1}^{L-1}(J_1{\tau}_{x,y}^+ {s}_{x+\frac{1}{2},y}^+{\tau}_{x+1,y}^- + J_2{\tau}_{x,y}^+ {s}_{x,y+\frac{1}{2}}^+{\tau}_{x,y+1}^- +\text{h.c.}).
\end{equation}
 Here we use $\tau$ to represent the spins on lattice sites and $s$ to represent the spins on lattice links. The open boundary condition is assumed. The local gauge generator is given by 
 \begin{equation}
     G_{x,y}=\tau_{x,y}^z-(s^z_{x+\frac{1}{2},y}-s^z_{x-\frac{1}{2},y})-(s^z_{x,y+\frac{1}{2}}-s^z_{x,y-\frac{1}{2}}),
 \end{equation}
with a suitable modification near the boundaries and the corners. 

Now we couple the dynamical gauge fields on lattice links to a Markov bath. The jump operators are still given by biased spin-flipping processes:
 \begin{equation}
 \begin{split}
     L_{x+\frac{1}{2},y}^{(u)}=\sqrt{\gamma_u}s_{x+\frac{1}{2},y}^+,\ L_{x+\frac{1}{2},y}^{(d)}=\sqrt{\gamma_d}s_{x+\frac{1}{2},y}^-,\\ L_{x,y+\frac{1}{2}}^{(u)}=\sqrt{\gamma_u^\prime}s_{x,y+\frac{1}{2}}^+,\ L_{x,y+\frac{1}{2}}^{(u)}=\sqrt{\gamma_d^\prime}s_{x,y+\frac{1}{2}}^-.
     \end{split}
 \end{equation}

The same symmetry-based construction method can be introduced in this 2D model. We consider the open quantum system $\mathcal{L}=\mathcal{L}_H+\mathcal{L}_D$, with the Hamiltonian and jump operators given above. 

When $\mathcal{L}_H=0$, the steady state behaves like $\rho=\rho_\tau\otimes\rho_s$ with $\rho_{s}=\otimes_{x.y}\rho_{s,(x+\frac{1}{2},y)}\otimes_{x.y}\rho_{s,(x,y+\frac{1}{2})}$. With only dissipative terms, we can easily find that $\rho_{s,(x+\frac{1}{2},y)}\propto \exp(\ln\beta s_{x+\frac{1}{2},y}^z)$ and  $\rho_{s,(x,y+\frac{1}{2})}\propto \exp(\ln\beta^\prime s_{x,y+\frac{1}{2}}^z)$ with $\beta=\gamma_u/\gamma_d$ and $\beta^\prime=\gamma_u^\prime/\gamma_d^\prime$. Therefore, we construct the double-space similarity transformation $\mathcal{T}_s$ acting on the link Hilbert space such that $\mathcal{T}_s\mathcal{L}_D\mathcal{T}_s^{-1}=\mathcal{L}_D^\dagger$. The double space transformation is $\mathcal{T}_s=T_s\otimes I_s$, where $T_s=\exp[-\sum_{x,y}(\ln\beta s_{x+\frac{1}{2},y}^z+\ln\beta^\prime s_{x,y+\frac{1}{2}}^z)]$ and $I_s$ is the identity matrix in the link Hilbert space.

The next job is to construct a double-space similarity transformation $\mathcal{T}=\mathcal{T}_\tau\otimes\mathcal{T}_s$ such that $\mathcal{T}\mathcal{L}_H\mathcal{T}^{-1}=\mathcal{L}_H$. With the above $\mathcal{T}_s$, the $\mathcal{T}=T\otimes I$ can be obtained by a generalized gauge transformation: $T=\exp[-\sum_{x,y}(\ln\alpha+x\ln\beta +y \ln \beta^\prime)G_{x,y}]$. With the definition of local gauge generators, it is straightforward to check that restricting $T$   to the link Hilbert space leads to the above $T_s$. As a result, the steady state of the 2D dissipative quantum link model takes the form
\begin{equation}
    \rho_{\text{ss}}\propto\exp\left[\sum_{x,y}(\ln\alpha+x\ln\beta +y \ln \beta^\prime)G_{x,y}\right].
\end{equation}

The free parameter $\alpha$ is related to the strong global U(1) symmetry: $N=\sum_{x,y}G_{x,y}=\sum_{x,y}\tau_{x,y}^z$. This exact steady state exhibits the 2D many-body non-Hermitian skin effect, with the localization behavior in two directions determined by $\beta$ and $\beta^\prime$, respectively. All the characteristic features of its one-dimensional counterpart discussed in the main text—such as symmetry-resolved steady states, robustness against symmetric disorder, and the accumulation of dipole moments—also apply to this exact many-body steady state in two dimensions.

Therefore, this 2D example not only demonstrates that the dissipative quantum link model can be engineered to realize the 2D many-body non-Hermitian skin effect, but also highlights the broad applicability of our exact construction method, which is not constrained by dimensionality.

\bibliography{ref_LGT}